\documentclass[aip,reprint]{revtex4-1}

\usepackage{mathtools}

\usepackage{enumitem}
\usepackage{xr-hyper}
\usepackage{hyperref}
        \hypersetup{ % for arXiv to compile correctly
           breaklinks=true,   % splits links across lines
           colorlinks=false,   % displays links as colored text instead of blocks
           pdfusetitle=true,  % \title and \author values into pdf metadata
        }

\usepackage{lineno}
\usepackage{siunitx}
\usepackage{amsmath}
\usepackage{amssymb}
\usepackage{graphicx}
\usepackage[T1]{fontenc}
\usepackage{comment}

\usepackage[utf8]{inputenc}
\begin{document}

\title{Frequency-modulated combs via on-chip field enhancement}
\vspace{2 cm}

% authors to be discussed, just a draft
\author{Urban Senica}
 \email{usenica@phys.ethz.ch}
\author{Alexander Dikopoltsev}
\author{Andres Forrer}
%\author{Tudor Olariu}
%\author{Paolo Micheletti}

\affiliation{
Quantum Optoelectronics Group, Institute of Quantum Electronics, ETH Z{\"u}rich, 8093 Z{\"u}rich, Switzerland
}

\author{Sara Cibella}
\author{Guido Torrioli}
\affiliation{Istituto di Fotonica e Nanotecnologie, CNR, Via del Fosso del Cavaliere 100, 00133 Rome, Italy}

\author{Mattias Beck}
\author{J{\'e}r{\^o}me  Faist}
\author{Giacomo Scalari}
 \email{scalari@phys.ethz.ch}

\affiliation{
Quantum Optoelectronics Group, Institute of Quantum Electronics, ETH Z{\"u}rich, 8093 Z{\"u}rich, Switzerland
}
%\author{Urban Senica$^{1,*}$, Andres Forrer$^{1}$, Tudor Olariu$^{1}$, Paolo Micheletti$^{1}$, Sara Cibella$^{2}$, Guido Torrioli$^{2}$, Mattias Beck$^{1}$, J{\'e}r{\^o}me Faist$^{1}$, Giacomo Scalari$^{1,*}$}
%\email{usenica@phys.ethz.ch}

%\affiliation{
%Istituto di Fotonica e Nanotecnologie, CNR, Via del Fosso del Cavaliere 100, 00133 Rome, Italy
%}

\date{\today}% It is always \today, today,
             %  but any date may be explicitly specified

\begin{abstract}
%\textbf{\textcolor{red}{ABSTRACT TO BE REWRITTEN!}}\par 
Frequency-modulated (FM) combs feature flat intensity spectra with a linear frequency chirp, useful for metrology and sensing applications. Generating FM combs in semiconductor lasers generally requires a fast saturable gain, usually limited by the intrinsic gain medium properties. Here, we show how a spatial modulation of the laser gain medium can enhance the gain saturation dynamics and nonlinearities to generate self-starting FM combs. We demonstrate this with tapered planarized THz quantum cascade lasers (QCLs). While simple ridge THz QCLs typically generate combs which are a mixture of amplitude and frequency modulation, the on-chip field enhancement resulting from extreme spatial confinement leads to an ultrafast saturable gain regime, generating a pure FM comb with a flatter intensity spectrum, a clear linear frequency chirp and very intense beatnotes up to -30 dBm. The observed linear frequency chirp is reproduced using a spatially inhomogeneous mean-field theory model which confirms the crucial role of field enhancement. In addition, the modified spatial temperature distribution within the waveguide results in an improved high-temperature comb operation, up to a heat sink temperature of 115 K, with comb bandwidths of 600 GHz at 90 K. The spatial inhomogeneity also leads to dynamic switching between various harmonic states in the same device.

\end{abstract}

%\setboolean{displaycopyright}{true}

% Force line breaks with \\

%\date{\today}% It is always \today, today,
             %  but any date may be explicitly specified
             %AUTHOR LIST TO DISCUSS/MODIFY. 
%Does Tabea contribute fundamentally to the measurements? otherwise we put her in Acknowledgments \\

\maketitle
 % INTRODUCTION: THz QCLs, frequency combs & planarized WG geometry

%\textbf{\textcolor{red}{TO DO: INTRODUCTION}}
% INTRODUCTION
% - THz QCLs & combs: advantages, key references
% - reference the planarized waveguide platform
% - limitations: limited operating temperature, non-flat-top spectrum (not so useful), sometimes low RF beatnotes (difficult to measure)
\section*{Introduction}
Terahertz (THz) quantum cascade lasers (QCLs) are compact sources of coherent THz radiation based on intersubband transitions in an engineered semiconductor superlattice heterostructure \cite{kohler_terahertz_2002}. Owing to their relatively fast gain saturation nonlinear properties, they can operate as frequency combs \cite{burghoff_terahertz_2014, rosch_octave-spanning_2015} and dual combs \cite{rosch_-chip_2016}. Along with their significantly higher output powers than THz TDS systems \cite{neu2018tutorial} for frequencies above $\approx$1.5 THz, these devices are appealing for use in broadband coherent spectroscopy and sensing.

Recent important milestones in THz QCL development include advances in high-temperature narrowband operation \cite{bosco_thermoelectrically_2019, khalatpour_high-power_2020, khalatpour2023enhanced}, comb formation in ring cavities \cite{jaidl2021comb, micheletti2023terahertz}, heterogeneous integration on silicon substrates \cite{jaidl2022silicon}, operation as fast detectors \cite{Micheletti2021,pistore2022self}, and the development of a planarized waveguide platform with improved dispersion, RF and thermal properties \cite{senica2022planarized}. 

% why FM combs would be better: flat spectrum -> easier to measure, parabolic phase profile (can be re-compressed)
% bit unclear how you shift from THz to mid-IR. I whould make the transition smoother
For use in spectroscopy, combs with flat intensity spectra are often desired, as this relaxes the conditions on the required signal-to-noise ratio (or integration time) for measuring all the spectral components. From this perspective, mid-IR QCLs are considered more suitable than THz QCLs due to their fast saturable gain, which is the main cause of self-starting frequency-modulated (FM) combs  \cite{opacak_theory_2019-1}. Besides producing a flat intensity spectrum, their linear frequency chirp and parabolic phase profile also make external pulse compression schemes relatively straightforward\cite{taschler2021femtosecond}. 

In THz QCLs, the longer upper state lifetimes produce free-running comb states which are a mixture of amplitude and frequency modulation\cite{burghoff_terahertz_2014}. The amplitudes of the individual lines in the comb spectrum can often vary significantly, and comb operation is also limited to relatively low operating temperatures.

In this work, we show that spatial variation along the cavity, specifically tapering the laser, can lead to an effective increase in the speed of gain saturation dynamics. We find that the spatial field enhancement leads to an ultrafast saturable gain regime which produces pure FM combs with a flatter intensity spectrum, a linear frequency chirp and strong measured RF beatnotes. Moreover, the spatial modulation of the cavity width improves the resilience to high temperatures and also enables switching between various harmonic comb states on a single device.

% TAPERED DEVICE
% - geometry: 80 um wider regions for gain, 20 um narrower for mode selection, in-between adiabatic linear tapers (see also the numerical calculations)
% - field enhancement: wide-to-narrow transition, in  the idealized case (neglect coupling losses and overlap factor reduction) x4 -> x4^3 (enhanced four-wave mixing), results in very strong; write some equations! 
% measurement results: free-running BNs, up to nearly -30 dBm, three orders of magnitude higher than for comparable ridge devices 
% - field enhancement expansion: narrow parts gain saturation -> nonlinearity originates from there (curvature of gain); refer to Fig. 1(c) -> increased optical field increases the photon-driven transport, effectively shortens the carrier lifetime -> explain in the FM comb part?

\begin{figure*}[htbp!]
\centering
\includegraphics[width=0.9\linewidth]{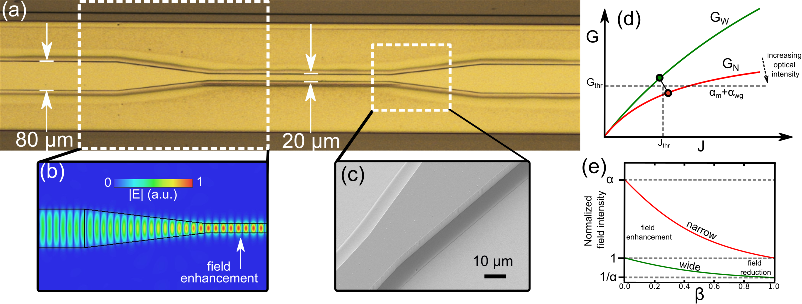}
\caption{\textbf{(a)} Optical microscope image of a tapered device, where wide (\SI{80}{\micro\metre}) and narrow (\SI{20}{\micro\metre}) sections are connected via adiabatic linear tapers. \textbf{(b)} The E-field distribution obtained from full-wave 3D numerical simulations reveals a strong field enhancement effect in the narrow taper sections. Assuming no scattering loss or overlap factor reduction, the field intensity enhancement is proportional to the width ratios, in this case 4:1. This results in an enhanced four-wave mixing nonlinear process, crucial for frequency comb generation. \textbf{(c)} SEM image of the tapered active waveguide after the dry etching process step with visible vertical, smooth sidewalls. Subsequently, the active waveguides are planarized with a low-loss polymer (BCB) and covered with an extended top metallization. \textbf{(d)} Illustration of the gain in the wide and narrow sections, including intensity-dependent gain saturation. In the narrow sections, due to stronger gain saturation and photon-driven transport, the gain is lower but more nonlinear. \textbf{(e)} Field intensity in the narrow and wide sections as a function of the width ratios $\alpha$ and the fraction of the device length with narrow sections $\beta$. The values are normalized to the field intensity of a homogeneous waveguide with the same properties and operation point. 
}
\label{fig:Taper_RF}
\end{figure*}

\section*{Tapered planarized waveguide}
\subsection*{Waveguide geometry}
We designed and fabricated tapered waveguides using a homogeneous broadband THz QCL active region \cite{forrer_photon-driven_2020} and our planarized waveguide platform \cite{senica2022planarized}. 
As shown in the optical microscope image in Fig. \ref{fig:Taper_RF}.(a), the tapered waveguide consists of a sequence of wide (\SI{80}{\micro\metre}) and narrow (\SI{20}{\micro\metre}) sections connected with adiabatic linear tapers that minimize scattering losses. 
While in the mid-IR, tapered active waveguide geometries have recently been shown to improve the frequency comb performance due to a lower overall chromatic dispersion \cite{Wang2020}, in our implementation there are several other crucial effects due to the tapered geometry.

The narrow sections act as a filter for selecting the fundamental transversal waveguide mode (required for comb operation with regular teeth \cite{bachmann_short_2016}), without using any side absorbers. The wider sections provide more gain for higher output power and a broader emission spectrum due to lower waveguide losses. While fabricating a homogeneous waveguide with the narrow width of \SI{20}{\micro\metre} would be beneficial for transversal mode selection and heat dissipation, the increased dispersion and waveguide losses would severely limit the total comb bandwidth and output power. The propagating mode simulations in Fig. \ref{fig:Taper_RF}.(b) show the efficient, scattering-free transition between the wide and narrow sections, which do not affect the formation of longitudinal modes and standing waves along the cavity length.

\subsection*{Field enhancement and nonlinearity}
Due to the non-homogeneous waveguide width, there is a field enhancement effect in the narrow sections, as shown in the simulation results in Fig. \ref{fig:Taper_RF}(b). Considering an idealized case (neglecting any reflection or scattering loss and reduction of the overlap factor $\Gamma$), the field intensity enhancement ($FIE$) is proportional to the width ratio between the wide and narrow sections (in this case 4:1), while the field amplitude enhancement ($FAE$) scales with the square root of the ratio ($\sqrt{4}$:1=2:1). % didn't we say the field is enhanced by 2 and the intensity/power is by 4? -> yes, writing for field intensity now
This is an important aspect, as the spontaneous frequency comb formation in QCLs is based on the nonlinear four-wave mixing process\cite{hugi_mid-infrared_2012}. Since the latter is a third-order process, its efficiency is proportional to the cube of the electric field intensity \cite{friedli2013four}. % We said that the efficiency grows like E^2 -> referring to field intensities!
For example, if we consider the case of non-degenerate four-wave mixing with two initial frequencies of $\omega_1$ and $\omega_2$, due to the $\chi^{(3)}$ (Kerr) nonlinearity within the active region waveguide \cite{opavcak2021frequency}, two new frequencies $\omega_3=2\omega_1-\omega_2$ and $\omega_4=2\omega_2-\omega_1$ will be generated with an output field intensity proportional to $I_3\propto |\chi^{(3)}|^2 I_1^2 I_2$ and $I_4\propto |\chi^{(3)}|^2 I_1 I_2^2$, % are you sure about this formula? check formula and ref -> good point, corrected to |Chi(3)|^2
where $I_i$ is the field intensity of the lasing mode at frequency $\omega_i$\cite{boyd_nonlinear_2008}. 
%Due to the field enhancement in the tapered waveguide, both the intensity of the input lasing modes $I_1, I_2$ and the nonlinearity $\chi^{(3)}$ are increased in the narrow waveguide sections, resulting in an increased efficiency of four wave mixing.
%For our specific geometry, taking the width ratios into account, this would result in a field intensity of the new modes generated by four-wave mixing to be increased by a factor of $4^3=64$ in an ideal case. % 64? what, are you crazy?

The nonlinearity within the active region arises mainly from gain saturation, i.e., the fact that the gain changes as a function of the intracavity optical field intensity \cite{indjin2002self}:
\begin{align}
\label{eq:gain_sat}
    g = \frac{g_0}{1+I/{I_\mathrm{sat}}}
\end{align}
where $g_0$ is the unsaturated gain, $I$ the intracavity optical field intensity, and $I_\mathrm{sat}$ the saturation intensity. 

With an increasing intracavity optical intensity, the gain will be reduced, while the nonlinearity will be enhanced (and vice versa).
%suggested by Jerome
Since the active region of a QCL is providing both gain and nonlinearity, a trade-off between these two quantities arises naturally in ridge devices with a constant width. Achieving a good figure of merit in both simultaneously is challenging, as the nonlinearity will be maximum in the highly saturated gain response. With the tapered waveguide, we can, however, use the best of both: the wide sections provide a larger gain, while the narrow sections provide an increased  nonlinearity, which is a regime generally not accessible without field enhancement.

This is illustrated in Fig. \ref{fig:Taper_RF}(c), where we plot the modal gain at an increasing current density through the active region. The dependence for the wide (G$_{\mathrm{W}}$) and narrow (G$_{\mathrm{N}}$) sections were computed using the relation from Eg. \ref{eq:gain_sat} with $I_{\mathrm{N}}=4 \times I_{\mathrm{W}}$, and with the assumption that the unsaturated gain $g_0$ is increasing linearly with the applied laser bias \cite{barbieri2000gain}. Due to the field enhancement effect, the gain in the narrow sections is decreased while the nonlinearity increases (larger curvature). To sustain lasing, the total gain of the cavity must overcome the total cavity losses (waveguide + mirror losses, as indicated by the gray horizontal line). The operating point is marked with the colored circles: due to the reduced gain in the narrow sections, the wider sections operate at a point with a higher gain.

% TO DO: write nicely
While the relative field intensities in the narrow sections are enhanced by the width ratio compared to the wider sections, it is the absolute field intensities that are crucial in nonlinear processes. For a homogeneous ridge active waveguide, the steady state (average) intracavity field intensity will increase with the length of the cavity and with the mirror reflectivities. 

We now compare the intracavity field intensities within a tapered and a ridge waveguide with several simplifying assumptions: both waveguides are made of the same active material, and have the same waveguide losses, mirror reflectivities and cavity length. The tapered waveguide is approximated as consisting of only the wide and narrow sections, neglecting the tapered transitions between them.
Under these assumptions, the threshold gain $g_{\mathrm{thr}}$ of both devices would be the same, and we can write:
\begin{align}
\label{eq:I_normalized}
    g_{\mathrm{thr}}=\underbrace{\frac{(1-\beta)g_0}{1+I_\mathrm{W}/{I_\mathrm{sat}}}}_{\text{wide sections}}+\underbrace{\frac{\beta g_0}{1+\alpha I_\mathrm{W}/{I_\mathrm{sat}}}}_{\text{narrow sections}} = \underbrace{\frac{g_0}{1+ I_{H}/{I_\mathrm{sat}}}}_{\text{homogeneous waveguide}}
\end{align}
Here, $\alpha$ is the width ratio between the two sections, and $\beta$ is the fraction of the narrow sections within the whole device length.

Following from Eq. \ref{eq:I_normalized}, in Fig. \ref{fig:Taper_RF}(e) we show a general plot of the intracavity field intensities within an active multi-section waveguide with saturable gain, consisting of two different widths. The intensities are normalized to a homogeneous ridge waveguide with the same properties and at the same operating point. Depending on the filling factor $\beta$, the normalized field enhancement in the narrow sections (red) varies between 1 and $\alpha$. In the wide sections (green), the field is reduced to normalized values between 1 down to $1/\alpha$. 
We should note that the curvature of the dependence on $\beta$ (and consequently the normalized field enhancement at a specific point) changes also with the operation point (laser bias). 

From this, we can draw some more general conclusions for designing tapered waveguides. For maximal field enhancement in the narrow parts, $\alpha$ should be large and $\beta$ small. For minimizing the field in the wider section, both $\alpha$ and $\beta$ should be large (a similar approach is used in tapered amplifiers, where the field intensity spreads out in the tapered sections to reduce gain saturation\cite{walpole1996semiconductor}). However, for nonlinear processes, the interaction length within the waveguide matters as well, so for a given application, the optimal $\beta$ will lie somewhere between 0 and 1. Moreover, in practical devices also $\alpha$ cannot be arbitrarily large. For example, in our THz QCL geometry, the narrow sections are limited by the increasing waveguide losses and reduction of the mode overlap factor, while the wider sections are limited by a worse thermal figure of merit. 

%%% MEASUREMENT RESULTS %%%
% 1 - Typical measured spectrum w/ BN
% flat-top comb spectrum + super-strong RF BN + high T
% drawback: increased waveguide losses (higher threshold), doesn't increase comb bandwidth directly (due to larger losses)
\section*{Results}
In the following, we present measurement results of a 4.2 mm long tapered device with a total of three wide (\SI{80}{\micro\metre}) and two narrow (\SI{20}{\micro\metre}) sections, with cleaved end facets. For this specific device, $\beta$=0.36, while $\alpha$=3.16 (extracted from numerical wave propagation simulations). The device was soldered on a copper submount with a custom RF-optimized PCB \cite{senica2022planarized} and mounted on a flow cryostat.

\subsection*{Measured THz spectrum and RF beatnote}
In Fig. \ref{fig:Taper_RF}(a-c), we show a typical THz spectrum and RF beatnote measurement which highlights several performance improvements of the tapered geometry. The measurement was done at a relatively high heat sink temperature of 90 K, and the comb spectrum spans around 600 GHz. In contrast to ridge devices, where the individual mode amplitudes are typically varying over several orders of magnitude, we observe a flatter comb spectrum, where the modes in the central $\sim$300 GHz of the comb are within a $\sim$10 dB intensity variation. The measured free-running RF beatnote power is nearly -30 dBm, which is almost three orders of magnitude higher than for ridge devices processed on the same chip in Ref. \cite{senica2022planarized}. This is due to contributions of the field enhancement effect and a larger total intracavity optical power (larger device area with wider sections). 
% insert equations - four-wave mixing

The dependence of the measured RF beatnote intensity on the field enhancement can easily be explained with the following expression: as the free-running RF beatnote is a direct measurement of the current modulation $\Delta I(t)$ at the mode spacing frequency $f_\mathrm{rep}$, its intensity is proportional to the sum of the product of the neighbouring modes' complex electric field amplitudes\cite{li_dynamics_2015}:
\begin{align}
    \Delta I(t) \propto \sum_{i} E_i\ E_{i+1}^*
\end{align}

%\begin{align}
%    \Delta I(t) \propto \big(E_1\ e^{i \omega_1 t}\ + E_2\ e^{i \omega_2 t}\big)^2 = E_1^{^2}+E_2^{^2}+E_1\ E_2^*\ e^{i \delta t}
%\end{align}
Here, $E_i$, $E_{i+1}^*$ are the electric field amplitudes of two neighbouring modes in the THz emission spectrum, which contribute to a measurable signal at $f_\mathrm{rep}$.
% difference frequency generation -> depends on the square of the amplitudes!
Typical measured beatnotes also display narrow linewidths, in the order of $\sim$1 kHz. 

For tapered devices at such elevated heat sink temperatures (90 K as opposed to the more standard 15-40 K \cite{mezzapesa2020terahertz, yang_terahertz_2016}), the comb bandwidth is reduced due to two main contributions: increased waveguide losses in the narrow sections (\SI{20}{\micro\metre}), and the decreased gain due to an increased heat sink temperature. This results in the absence of the low frequency part of the emission spectrum for this epilayer (down to $\sim$2.4 THz). 

However, at lower heat sink temperatures (40 K), a comparable comb bandwidth to those obtained with simple planarized ridges with a width of \SI{40}{\micro\metre} is observed (typically spanning 700-800 GHz at 40 K). THz emission and RF beatnote spectra for a varying laser bias are shown in Figs. S1, S2 of the Supplemental Document, along with a movie of a laser bias sweep which displays a rich landscape of fundamental and harmonic comb states.
Additionally, a comparison of measured comb spectra of ridge and tapered devices is in Fig. S3, where it is evident that the tapered devices feature a significantly flatter envelope of the spectrum intensity.

\begin{figure}[htbp]
\centering
\includegraphics[width=1\linewidth]{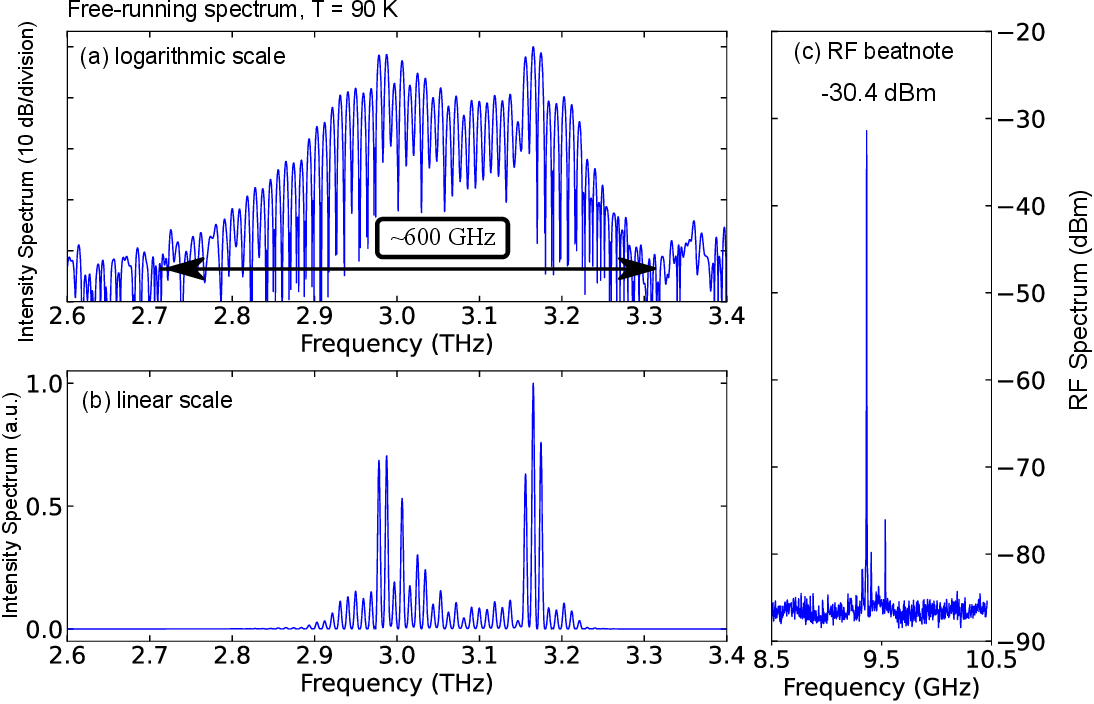}
\caption{\textbf{(a)} Measured THz emission spectrum in logarithmic scale with a comb bandwidth of around 600 GHz at a heat sink temperature of 90 K. \textbf{(b)} The same measured spectrum in linear scale displays a relatively flat comb spectrum between 2.9-3.2 THz. \textbf{(c)} Measured RF beatnote at the roundtrip frequency, up to nearly -30 dBm.
}
\label{fig:Spectrum_BN}
\end{figure}

\subsection*{Linear chirp and flatter spectrum}

% 4 - SWIFTS, linear chirp observed -> include also instantaneous frequency, looks similar to mid-IR QCLs, sth to add here?

\begin{figure*}[htbp!]
\centering
\includegraphics[width=1\linewidth]{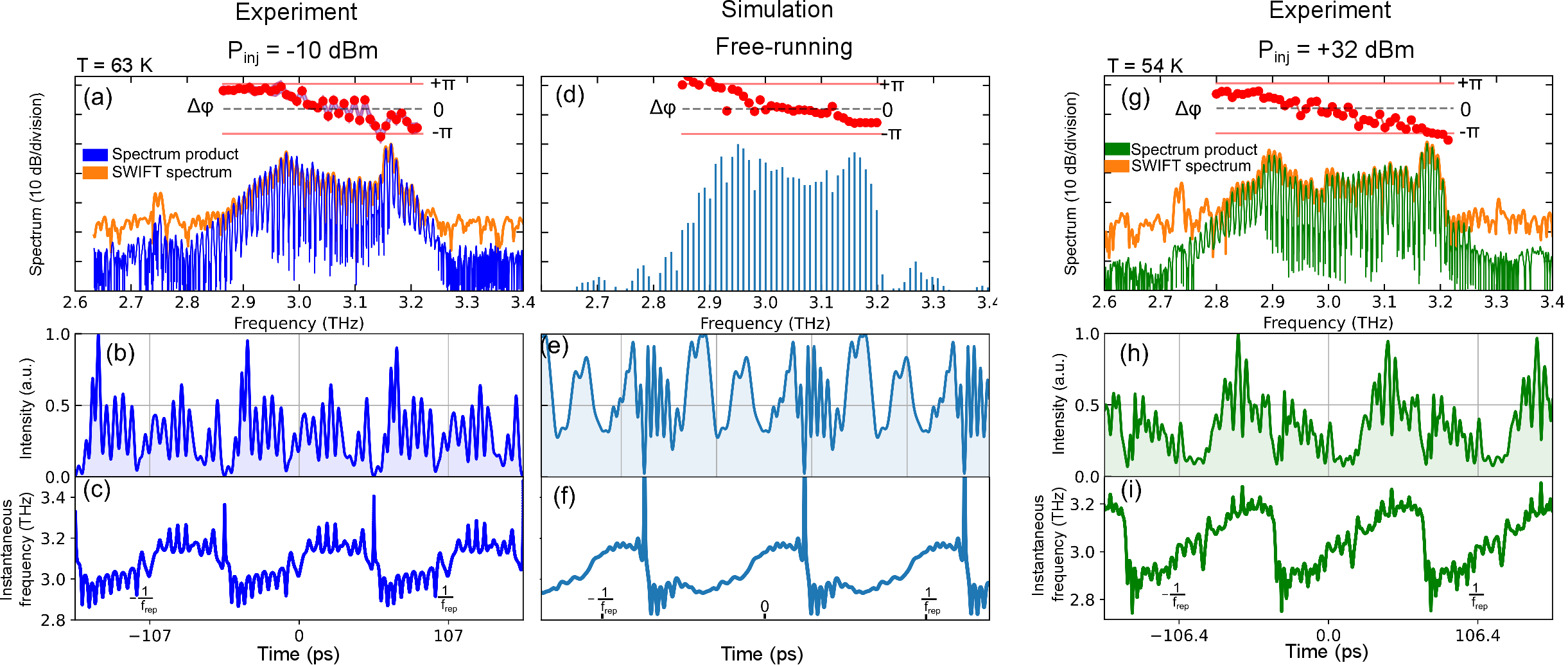}
\caption{SWIFT spectroscopy measurements and mean-field theory simulations. \textbf{(a-c)} Measurements on a weakly RF-injected (-10 dBm) tapered device produce a relatively flat comb emission spectrum, a linear frequency chirp and a quasi-continous output intensity with some oscillations. \textbf{(d-f)} Results of mean-field theory simulations with a spatial dependence of the crucial parameters are able to reproduce the main features of the measured device.
\textbf{(g-i)} Measurement results of the same tapered device under strong RF-injection (+32 dBm) with a broadened comb emission spectrum and a cleaner linear frequency chirp with a more constant output intensity on top of a sine wave which follows the RF modulation.
} 
\label{fig:swifts}
\end{figure*}

We then performed Shifted Wave Interference Fourier Transform (SWIFT) spectroscopy measurements \cite{burghoff_terahertz_2014, Burghoff2020} to assess the comb coherence and reconstruct the time domain profile using a fast hot electron bolometer (HEB) detector \cite{Semenov_2002, torrioli2023thz} (a more detailed explanation of the working principle and setup used is in Ref. \cite{senica2022planarized}). A relatively weak RF signal (-10 dBm at the RF source) was injected at the roundtrip frequency $f_\mathrm{rep}$ to stabilize the comb repetition rate and to give the QCL and the spectrum analyzer a common time-base allowing the IQ demodulation. For such a weak RF injection power it is assumed that the comb is stabilized without perturbing its free-running state (namely, the specific intensity spectrum and intermodal phases). In Fig. \ref{fig:swifts}(a), we plot both the spectrum product and the SWIFT spectrum. The first was obtained by measuring the DC interferogram with a slow detector (DTGS), while the latter was reconstructed with IQ demodulation from optical beatnote measurements with a fast detector (HEB). The excellent agreement and comparable signal-to-noise ratio is an indicator of good comb coherence, while also the detected optical beatnote measured on the HEB is in the order of 20-30 dB stronger than for planarized ridge samples. The reconstructed intermodal phase profile in this state follows a linear chirp, which is typically observed in mid-IR QCLs \cite{singleton_evidence_2018, cappelli2019retrieval}, but has not yet been reported so clearly in THz QCLs. 
In THz QCL designs optimized for comb operation, the longer upper state lifetime ($\tau_{\mathrm{up}}>10$ ps) and consequently the larger $\omega_\mathrm{rep}\times \tau_{\mathrm{up}}$ product makes the comb state less explicitly FM (see time reconstructions in \cite{burghoff2015evaluatingOPEX,cappelli2019retrieval}) compared to mid-IR QCL combs, where ultrafast gain saturation and gain asymmetry (arising from non-parabolicity and Bloch gain \cite{opavcak2021frequency}) play a major role in driving the laser dynamics. 

The reconstructed time profile has a quasi-continuous periodic output intensity with some oscillations, and the instantaneous frequency produces a linear chirp, both are shown in Fig. \ref{fig:swifts}(b, c). This could indicate that the effects related to the field enhancement push the tapered THz QCL comb towards a regime similar to mid-IR QCL combs, with a flatter intensity spectrum and a linear chirp in frequency. 

Indeed, another aspect of the field enhancement is the increased photon-driven current \cite{choi2008gain} in the narrow sections, which shortens the stimulated carrier lifetime $\tau_{\mathrm{st}}$ through the dependence:
\begin{align}
\label{eq:t_stim}
    \tau_{\mathrm{st}} = \frac{1}{g_{\mathrm{c}}\ S}
\end{align}
where $g_{\mathrm{c}}$ is the gain cross-section and $S$ the photon density. This results in a reduction of the upper state lifetime $\tau_{\mathrm{up}}^{-1}=\tau_{\mathrm{nr}}^{-1}+\tau_{\mathrm{sp}}^{-1}+\tau_{\mathrm{st}}^{-1}$, where $\tau_{\mathrm{nr}}, \tau_{\mathrm{sp}}$ are the non-radiative and spontaneous emission lifetimes, respectively. With this effective lifetime shortening, the tapered waveguide system goes into a fast saturable gain regime, where the laser tends to produce a continuous waveform with a quasi-constant optical intensity, manifested as frequency-modulated combs usually observed in mid-IR QCLs\cite{opacak_theory_2019-1}.

% we developed a model, and then simulated with a mean-field approach
% the resutls shown in fig. 1 have a good agreement with experiments
% note that we use the actual device parameters given in the supplementary
% we find that both the dependence of the gain and nonlinearity are important for this specific comb state
To study the mechanism driving the comb dynamics, we developed a model which includes a spatial dependence of the optical nonlinearities, gain saturation, and temperature distribution within the cavity (by modifying the gain profile), and performed numerical simulations following a mean-field theory approach based on Ref. \cite{burghoff2020unraveling}. The details of the model as well as all the simulation parameters used (Table S1) along with the frequency-dependent waveguide and material dispersion (Fig. S5) can be found in the Supplemental Document. Results obtained with our model have good agreement with the experimental results. As shown in Fig. \ref{fig:swifts}(d-f), the simulation produces a relatively flat spectrum separated into two main lobes with a phase distribution similar to the measured result. The linear frequency chirp is also reproduced in simulation, with a discontinuity at the phase jump point. The time domain profile reconstructs the quasi-continuous intensity with oscillations and individual points in time where the intensity almost vanishes, again consistent with the measurements. From the simulation model we found that the spatial dependence of the nonlinearities and gain are crucial for obtaining these specific kinds of comb states.

Sucessively, when increasing the RF injection power up to +32 dBm, the comb spectrum is further broadened and flattened, and the observed linear chirp has a cleaner shape, as visible in Fig. \ref{fig:swifts}(g-i). The THz intensity spectrum features a flatter profile as well, while its time domain intensity profile is modulated in amplitude, following the strong RF-injected cosine profile ($\propto$ cos$(2\pi f_\mathrm{rep} \ t)$). These observations are similar to the findings of extensive mid-IR QCL simulations reported in Ref. \cite{opavcak2022origin}. There, due to a non-zero third-order dispersion in the mid-IR QCL cavity, the intensity spectrum features oscillations in amplitude and the intermodal phases form groups, diverging from an ideal linear chirp (which is similar to some states observed in free-running THz QCLs). By means of RF injection, a clean linear chirp and flatter spectral amplitudes can be recovered, which is consistent also with our experimental results with the tapered planarized waveguide geometry.

%A more detailed explanation of the implemented mean-field theory simulation model and a complete list of parameters used can be found in the Supplemental Document, as well as the computed frequency-dependent material and waveguide dispersion of the tapered cavity.

\subsection*{High-temperature performance}
% 2 - high temp.: due to the enhanced four-wave mixing, comb operation with strong RF beatnotes observed almost to the max lasing temperature (usually it goes to single mode way before), up to even 115K in CW (record?). The lasing Tmax is large as well, as the wider sections are separated by narrower ones (acting as heatsinks)
\par
\begin{figure}[htbp]
\centering
\includegraphics[width=1\linewidth]{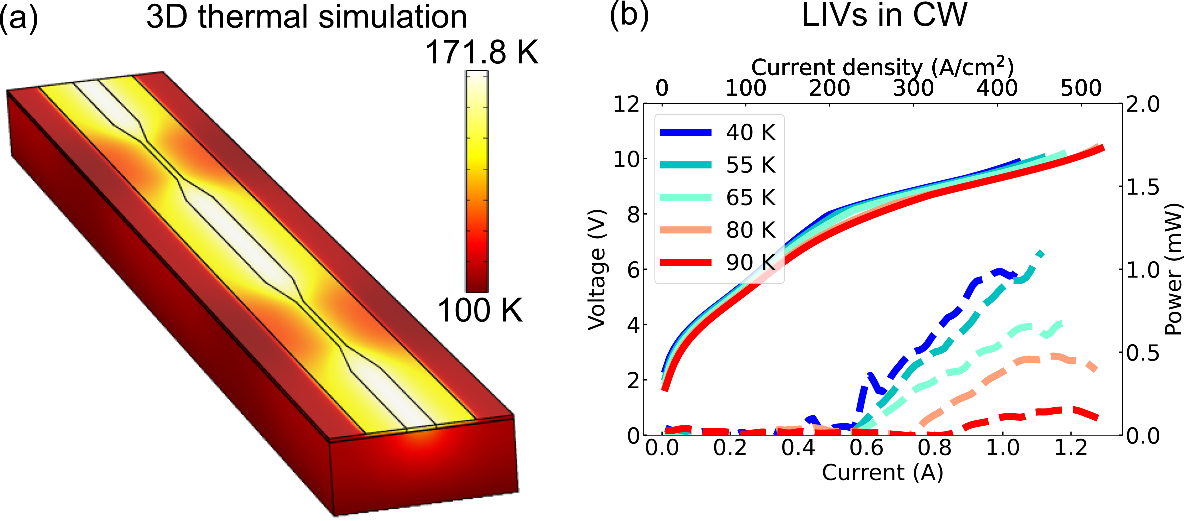}
\caption{\textbf{(a)} 3D thermal COMSOL simulations of the tapered device at the maximum laser bias and a heat sink temperature of 100 K. While the wider sections are heating up more, they are separated into smaller islands where the narrower sections act as heat dissipation channels with a lower operating temperature. \textbf{(b)} High-temperature LIV characteristics of a tapered device in CW. 
}
\label{fig:Taper_thermal}
\end{figure}

 Improved frequency comb properties are maintained for even higher operating temperatures. Results of a 3D COMSOL thermal simulation in Fig. \ref{fig:Taper_thermal}(a) show that the wider sections, which are heating up more, are separated into smaller islands and the connecting narrower regions act as heat dissipation channels. 
%This is more easily seen if we plot the temperature distribution along the device, see Fig. \ref{fig:Taper_thermal}(b). As a comparison to the tapered waveguide (blue line), the homogeneous temperature distribution  along a ridge waveguide with a constant width of \SI{40}{\micro\metre} is also shown (magenta line). 
In the thermal simulations, typical maximum bias conditions (11 V, 400 A/cm\textsuperscript{2}), a heat sink temperature of 100 K, and the following material heat conductivities were used: Cu = 320 W/mK, GaAs/AlGaAs active region = 5 W/mK \cite{Scamarcio2008}, GaAs substrate = 100 W/mK, BCB = 0.15 W/mK (half of the reported room temperature conductivity) \cite{Choy1977, Greig1988}.

In Fig. \ref{fig:Taper_thermal}(b), we plot LIV curves measured in continuous wave (CW) operation. The power measurements were done with a large area Thomas Keating absolute THz power meter and a chopper wheel, with limited sensitivity and without any correction for collection losses, so they could not be performed up to the maximum lasing temperature. Measured with a room-temperature DTGS detector and an FTIR, the maximum lasing temperature in CW was as high as 118 K.

\begin{figure*}[htbp]
\centering
\includegraphics[width=1\linewidth]{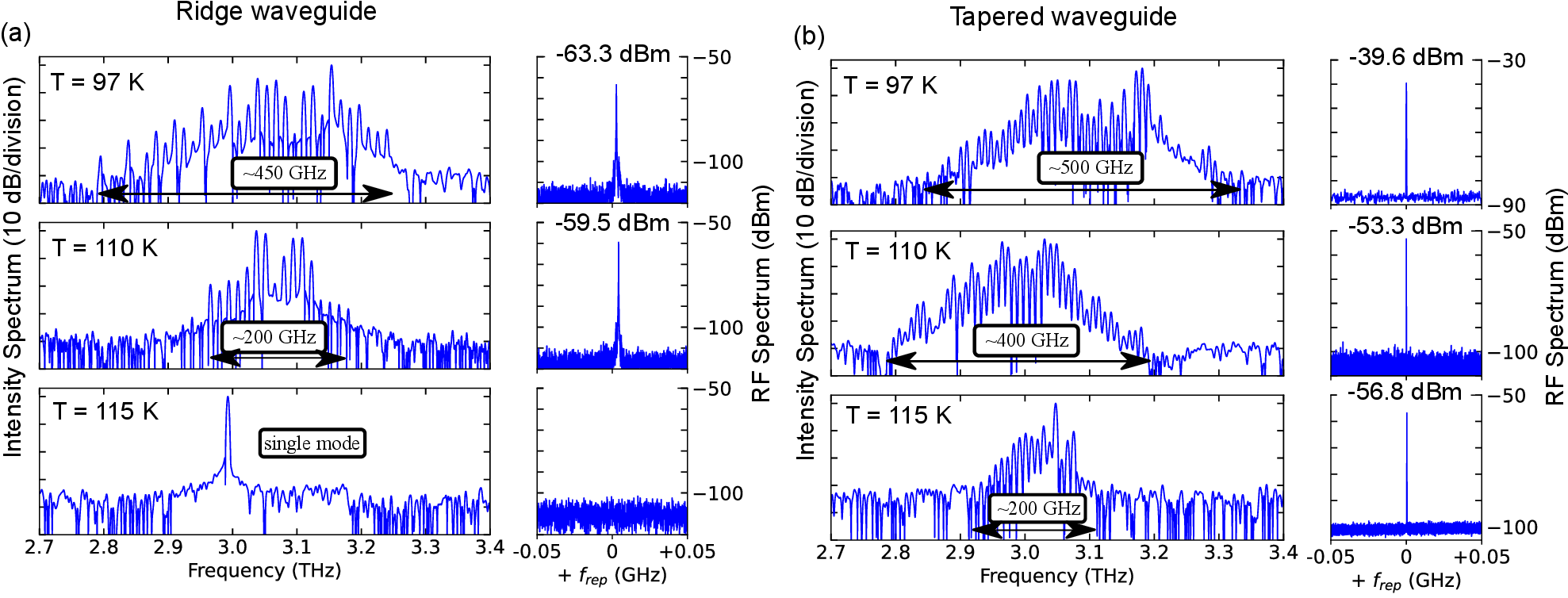}
\caption{Measured spectra and free-running RF beatnotes at high operating temperatures, comparing \textbf{(a)} a  ridge waveguide (W = \SI{40}{\micro\metre}, L = 2.7 mm) and \textbf{(b)} a tapered waveguide device (W = 80/20 \SI{}{\micro\metre}, L = 4.2 mm). The tapered waveguide device displays broader THz bandwidths, flatter emission spectra, stronger RF beatnotes, and a higher maximum comb temperature of 115 K. The spectra were obtained with a room-temperature DTGS detector.}
\label{fig:Comparison_thermal}
\end{figure*}

In Fig. \ref{fig:Comparison_thermal}(a, b) we compare the high-temperature comb operation of a tapered waveguide and a reference ridge waveguide, both fabricated during the same process run on the same chip. The ridge sample has a length of 2.7 mm and a constant waveguide width  of \SI{40}{\micro\metre}, with a maximum lasing temperature in CW up to 116.5 K. Comparing the two samples at the same heat sink temperatures, it can be seen that the tapered waveguide device features a broader comb bandwidth, a flatter THz emission spectrum, and a stronger RF beatnote. Comb operation is maintained up to 115 K, still with a bandwidth of  $\sim$200 GHz with a strong single RF beatnote above -60 dBm. This improved high-temperature comb performance is attributed to two main contributions. The first is the field enhancement effect giving rise to stronger nonlinearities and gain saturation which affect the comb formation process. Secondly, the narrow sections are at a significantly lower temperature than the wider regions (simulation results indicate a difference of $\sim$25 degrees). 

A more detailed thermal simulation result analysis can be found in Fig. S4 of the Supplemental Document, where the line cuts across different profiles of the device show a complex temperature distribution due to the non-homogeneous geometry consisting of materials with different thermal properties. %The lower temperature of the narrow tapered sections compared to the homogeneous ridge waveguide is also shown.

We should also note here that the observed threshold current density of around 180 A/cm\textsuperscript{2} at 40 K in CW of this device is higher than the ones we reported for simple planarized ridges with a width of \SI{40}{\micro\metre} (140 A/cm\textsuperscript{2} at 40 K), which is due to increased waveguide losses and gain saturation in the narrow sections with a width of only \SI{20}{\micro\metre}.

\subsection*{Harmonic comb state switching}
% 3 - harmonic comb engineering: due to the non-uniform shape, prone to switch to harmonic states; and it's possible for a single device to switch between various harmonic states just by varying the laser bias (with some hysteresis effects) -> we also record strong single BNs at the corresponding harmonic; use??
\par
\begin{figure}[htbp]
\centering
\includegraphics[width=1\linewidth]{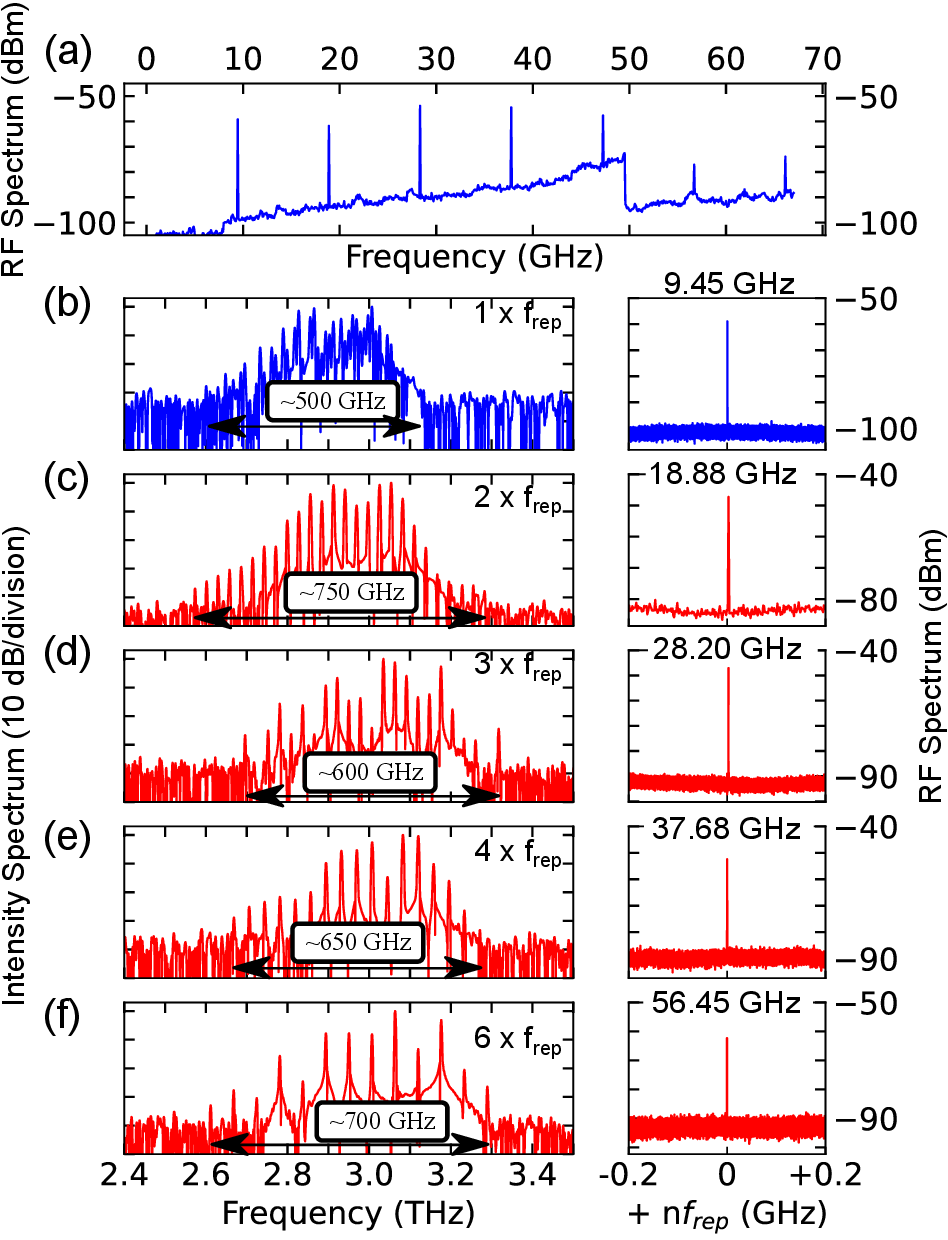}
\caption{Harmonic comb state switching: on a single tapered device, various pure harmonic comb states can be excited by varying the laser bias and temperature. \textbf{(a)} RF spectrum measured on a fundamental comb state from panel \textbf{(b)} shows strong RF signals at the fundamental and multiples of the repetition rate f\textsubscript{rep}, up to the 7\textsuperscript{th} harmonic (limited by the spectrum analyzer range). \textbf{(b-f)} highlights the measured spectra and RF beatnotes of the pure fundamental (blue) and harmonic comb states (red). All of the measurements were performed at a heat sink temperature of 80 K or higher, and there are no other RF signals present besides the strong single beatnote.}
\label{fig:harmonics}
\end{figure}

 Another unique feature of the tapered geometry is the possibility to switch to various harmonic comb states on demand. Recently, such harmonic comb states, where the lasing mode spacing is an integer of the fundamental f\textsubscript{rep}, have gained interest in the QCL community \cite{kazakov_self-starting_2017}. In mid-IR QCLs, engineered defects have been fabricated for selecting a specific harmonic order\cite{Kazakov2021}, while in THz these have been so far limited to spontaneously forming harmonic states without an external control \cite{ForrerAPL2021Harmonic, wang_harmonic_2020}. As our tapered devices have a non-uniform shape along the length of the device, they are prone to switch to harmonic comb states. We demonstrate harmonic comb state switching, where a plethora of pure harmonic comb states can be excited simply by varying the bias and temperature on a single tapered waveguide device. If we look at a fundamental comb state in Fig. \ref{fig:harmonics}(a, b), we can measure a strong RF beatnote at the fundamental f\textsubscript{rep}, but also at higher harmonics up to the 7\textsuperscript{th} harmonic (limited by the spectrum analyzer bandwidth of 67 GHz). This is an indication of strong comb coherence (maintained also between more distant comb lines). By varying the laser bias and temperature, we can switch between the fundamental and the 2\textsuperscript{nd}, 3\textsuperscript{rd}, 4\textsuperscript{th} and 6\textsuperscript{th} harmonic states, as shown in Fig. \ref{fig:harmonics}(b-f), respectively. These are pure harmonic comb states, as we detect a single RF beatnote only at the frequency of the harmonic mode spacing, without any other signals present in the RF spectrum. 

An interesting aspect is that the THz emission bandwidth of harmonic comb states is typically larger than for fundamental comb states, covering $\sim$700 GHz for the 6\textsuperscript{th} harmonic in Fig. \ref{fig:harmonics}(f) even at elevated heat sink temperatures above 80 K. In contrast to fundamental states, where also multi-beatnote or incoherent states are observed, harmonic states appear almost exclusively in a pure comb state. The harmonic combs in tapered devices produce strong RF beatnotes also for very high RF frequencies, still with a measured intensity of nearly -60 dBm at 56 GHz (we should note that this is without any correction for increased high-frequency cable and PCB losses). This feature makes them appealing also as reliable coherent sources of high-frequency RF signals or even millimetre waves \cite{kazakov_self-starting_2017, pistore2021millimeter}.

In the Supplemental Material, we added some more simulation details, including an analysis of 3D simulation results of the RF-field distribution at the corresponding harmonic microwave resonances (Figs. S6, S7), along with some additional explanation as to why the 5\textsuperscript{th} and 7\textsuperscript{th} harmonic comb states were not observed experimentally on this specific tapered device.

% conclusions: field enhancement for high T, flat-top spectrum, strong RF, harmonic comb engineering
% outlook: engineering for harmonic combs -> larger total bandwidth, maybe without increasing waveguide losses
%\textbf{\textcolor{red}{TO DO: CONCLUSIONS}}
\section*{Conclusion}
In conclusion, we have presented a method to engineer the comb states in THz QCLs by spatially modulating the transverse dimension of a Fabry-P\'erot resonator. Such geometries are enabled by the planarized waveguide platform. A strong field enhancement effect results in an enhanced four-wave mixing process and gain saturation, both crucial for comb formation in THz QCLs. Measured devices produce flatter comb spectra spanning 600 GHz at a heatsink temperature of 90 K, with strong RF beatnotes up to nearly -30 dBm. Improved comb properties are maintained for high operating temperatures, with a comb bandwidth of 200 GHz at 115 K. We also report on the first experimental observation of a clear linear frequency chirp in a THz QCL, due to the device operating in a fast saturable gain regime. We are able to reproduce the results with a mean-field theory simulation model with a spatial dependence of the optical nonlinearities, gain saturation, and active region temperature.

In a broader context, electromagnetic environment engineering strongly affects light-matter interaction. One may consider our experiments as an analog to the Purcell effect, where spontaneous emission can be enhanced by light confinement in a cavity \cite{gu2013purcell}. In our work, the presence of strong subwavelength confinement with field enhancement leads to an ultrafast stimulated emission lifetime affecting the laser dynamics. 

It is important to emphasize that the presented field enhancement approach is not limited to THz QCLs, but can be applied to a variety of other laser systems\cite{hillbrand2020phase}, where a modified photon flux can be used to change the upper state lifetime and affect the laser dynamics by changing the value of the $\omega_\mathrm{rep}\times \tau_{\mathrm{up}}$ product. For example, in spectroscopy, a flatter comb spectrum is desired, and this can be produced with FM combs via strong field enhancement, as demonstrated in this work ($\omega_\mathrm{rep}\times \tau_{\mathrm{up}}$ is small). On the other hand, an increased waveguide cross-section would reduce gain saturation leading to slower dynamics, possibly facilitating short pulse formation in the presence of an increased $\omega_\mathrm{rep}\times \tau_{\mathrm{up}}$ product.

The tapered geometry also enables the switching between various pure harmonic comb states with increased comb bandwidths up to 750 GHz above 80 K. With the planarized waveguide platform, it should also be possible to engineer the switching to harmonic comb states by, for example, designing only the extended top metallization to match the shape of the harmonic RF field (by following the position of nodes and antinodes), without increasing waveguide losses as is the case in the tapered active waveguide geometry. This should in turn lead to increased absolute harmonic comb bandwidths. 

Beyond improved frequency comb performance, such integrated field enhancement structures can be used to boost the nonlinear optical properties originating from the large $\chi^{(2)}$ and $\chi^{(3)}$ nonlinearities within the active region, which could lead to novel effects and functionalities, such as the generation of new frequencies (e.g., via difference frequency \cite{belkin2007terahertz} and/or harmonic generation \cite{gmachl2003optimized}). 

%\newpage 

\ % The empty page
\begin{comment}
\subsection*{Methods}\label{method}
\textbf{\small{Planarized tapered waveguide fabrication}}

The active material tapered waveguides were fabricated following a standard double metal waveguide process (the main steps include wafer bonding, substrate removal, lithography with top metal evaporation and lift-off, and ICP dry etching). Subsequently, the structures were planarized with a low-loss polymer BCB (Cyclotene 3022-57) and an extended top metallization was deposited, spanning over the active and BCB-covered areas on the sides for placing the bonding wires. 
The full planarized waveguide fabrication details can be found in Ref. \cite{senica2022planarized}.
\vspace{0.5 cm}

\textbf{\small{Electromagnetic and thermal simulations}}

The electromagnetic 3D propagation and 2D eigenmode simulations were performed with full-wave solvers in the CST Microwave Studio and COMSOL simulation environments, respectively. The 3D thermal dissipation simulations were performed in COMSOL with the Heat Transfer module. More details are in the Supplemental Document.
\vspace{0.5 cm}

\textbf{\small{Spatially inhomogeneous mean-field theory model}}

The spatially inhomogeneous model was simulated numerically following a mean-field theory approach based on Ref. \cite{burghoff2020unraveling}. More details on the derivation and the full table of parameters used can be found in the Supplemental Document.

\end{comment}

\subsection*{Acknowledgements} The authors gratefully acknowledge funding from the ERC Grant CHIC (No. 724344) and in part by the Innosuisse (grant 53098.1 IP-ENG), and Actphast 4 Researchers P2020-41.
 %We also thank  Tabea Bühler, Sebastian Gloor and Johannes Hillbrand for technical help.
\subsection*{Competing Interests} The authors declare that they have no competing financial interests.

\subsection*{Authors contributions} U.S. and G.S. conceived the idea. U.S. designed and fabricated the devices, carried out all the measurements, analysed experimental data and performed electromagnetic and thermal numerical simulations under the supervision of G.S. and J.F. A.D. developed and implemented the spatially inhomogeneous mean-field theory model. A. F. built the SWIFTS setup. S.C. and G.T. provided the HEB detectors, A.F. and G.S. optimized the HEB RF coupling. M.B. performed the epitaxial growth. U.S. and G.S. wrote the manuscript. All authors discussed the results and commented on the manuscript.
 
\subsection*{Correspondence}  *Correspondence should be addressed to U. Senica (email: usenica@phys.ethz.ch) and G. Scalari (email: scalari@phys.ethz.ch).

\subsection*{Data availability}  All the simulation and experimental data supporting this study are available from the corresponding author upon reasonable request.

\subsection*{Keywords}  frequency combs, frequency modulation, field enhancement, terahertz, quantum cascade lasers
%\end{addendum}

%\end{addendum}

\section*{References}\label{References}
\bibliography{GS_bib_PlanarizedWG.bib}
% Produces the bibliography via BibTeX.

\end{document}